\documentclass[pdflatex,12pt]{article}
\usepackage[margin=1.2in,letterpaper]{geometry}
\usepackage{graphicx,subfigure,color}
\usepackage{amsmath,amsfonts,amssymb}
\usepackage{physics,slashed}
\usepackage{here, comment,hyperref,tcolorbox}
\usepackage{natbib}
\bibpunct[:]{[}{]}{,}{n}{}{,}

\title{
	First order phase transition  in the D3-D7 model  from the point of view of the  fermionic spectral functions
	}
\author{\small  
	Xian-Hui Ge$^a$\thanks{%
	E-mail:
	\href{mailto:gexh@shu.edu.cn}{gexh@shu.edu.cn}
	},~
	Shuta Ishigaki$^a$\thanks{%
	E-mail:
	\href{mailto:shutaishigaki@shu.edu.cn}{shutaishigaki@shu.edu.cn}
	},~
	Sang-Jin Sin$^b$\thanks{%
	E-mail:
	\href{mailto:sangjin.sin@gmail.com}{sangjin.sin@gmail.com}
	},
	and
	Taewon Yuk$^b$\thanks{%
	E-mail:
	\href{mailto:tae1yuk@gmail.com}{tae1yuk@gmail.com}
	}
	\vspace{5pt}
}
\date{\small%
	$^a${\it Department of Physics, Shanghai University, Shanghai 200444, China} \\
	$^b${\it Department of Physics, Hanyang University, Seoul 04763, Korea}
}
\begin{document}
\maketitle
\begin{abstract}
We consider the D3-D7 model and use  the  spectral function  of a probe fermion on D7 
  to analyze  the first order phase transition from the black-hole embedding phase to another black-hole embedding phase   in the presence of the  finite density and temperature. From the fermionic spectral functions, we study the temperature dependence of the decay rate, and we observe various phenomena that support the first order phase transition  including jump in it at the critical temperature that corresponds to the first order phase transition.
  %We found that if we assume that the resistivity is proportional to the decay rate as in the case of experiment or Drude model,  then the jump matches the resistivity  data in a recent  heavy fermion material.  
\end{abstract} 
\newpage
\tableofcontents

\section{Introduction}\label{sec:intro}
Shortly after the AdS/CFT correspondence \cite{Maldacena1997,Gubser1998,Witten1998} was established, the method has been applied to investigate quantum chromodynamics (QCD) and condensed matter physics of strongly correlated systems widely.
One of the benefits of the holographic method is that it allows us to  study strongly coupled quantum many-particles systems easily even for the system  with finite temperature and finite density effects.

The D3-D7 model \cite{Karch:2002sh} is one of the top-down models that has been used widely for such a purpose.
The background Schwarzschild-AdS$_5\times\mathrm{S}^5$ spacetime generated from the D3-branes and the probe D7-brane play roles of the thermal reservoir and charged particle system, respectively.
The solutions and behavior of the probe brane in the black hole spacetime, i.e., finite-temperature cases, were studied by refs.~\cite{Nakamura:2006xk,Nakamura:2007nx,Mateos:2006nu, Mateos:2007vn}.
The system has intricate dependence on the parameters exhibiting the phase transitions.
One of the solutions that the probe brane falls into the black-hole horizon, called black-hole embedding, is interpreted as a deconfinement phase of the quarks or metallic phase of the electrons.
The authors of Refs.\ \cite{Nakamura:2006xk,Nakamura:2007nx}
found that  there are two different phases of the brane embeddings where the D7-brane touches the black-hole horizon, and, as the density increases, there is a jump in the position of the horizon touching point. 
The brane in a phase named the BH-I phase bends sharper than the brane in another phase named the BH-II phase. 
The difference of these phases should have a physical interpretation from the viewpoint of the boundary field theory.
Since the black-hole touching configurations should be related to the metallic phase, the phase transition should be related to the metal-to-metal phase transition.
Therefore we can expect that certain phase transition in spectrum or transport property.
However, the details of the physical meaning of the two phases has been completely obscure. 

%Spectral functions can provide us with information on the system including the transport properties.
Although the spectral functions of bosonic fluctuations in this model were studied in Refs.\ \cite{Kruczenski:2003be,Erdmenger:2007ja,Myers:2007we,Mas:2008jz}, the fermionic spectral functions are a more interesting quantity, because they can be directly measured  by angle-resolved photoemission spectroscopy (ARPES) experiments.
%Since we are interested in the deconfinement phase of the quark, we can expect to measure the spectral function of the quark.
%Of course, however, it is impossible to compute the quark operator itself in the framework of the AdS/CFT corresponding because it is not color singlet.
In this work, therefore, we introduce  a probe fermion field of bottom up character  living on  the D7 model  and compute the fermionic spectral function. 
%We expect that the fermion can qualitatively mimic the electron in some situation and the  the deconfined phase describe a thermalized electron state.  The  fermion spectral functions for the bottom up model  were already studied extensively. 
%The authors of refs.~\cite{Liu:2009dm,Faulkner:2009wj} studied a holographic model with a fermionic field using the methodology of  \cite{Iqbal:2009fd}, and suggested  that the model represent a class of non-Fermi liquid of strongly correlated electron system. 
%This  bottom-up model seems to have a Fermi surface but many  properties are different from those of the Fermi liquid.
%The authors of ref.~\cite{Faulkner:2009am} studied a bottom-up model with a coupling between the fermionic fields and the scalar condensation and tried to mimic a gap in the fermionic spectral function.
%Recently,  similar bottom-up models has been studied by refs.~\cite{Oh:2021xbe,Oh:2020cym} and in ref.~\cite{Yuk:2022lof},  the authors   demonstrated  the similarities of gap structure  between the holographic model and the BCS model.
The  fermion spectral functions for the bottom-up model  were  already studied extensively \cite{Liu:2009dm,Faulkner:2009wj,Iqbal:2009fd, Faulkner:2009am,Oh:2021xbe,Oh:2020cym, Yuk:2022lof}.
However, the fermion spectral function in fundamental representation with the D3-D7 model has not been studied much.
%One of the motivations of this paper is to help understanding of the strongly correlated electron system by considering fermion spectral function with top down rather than bottom up model. 
% relationship between such holographic models and high-$T_c$ superconductors from a new setup of the holographic model with a fermionic field.The fermionic spectral function can be obtained by analyzing the bulk fermionic field in the holographic method.
Since we want the fermion to  provide the  matter density,  the   natural candidate of the fermionic field in the D7-brane is the fermionic degree of freedom coming from the D3-D7 string \cite{Mateos:2007vc} rather than 
 the mesino   \cite{Martucci:2005rb,Kirsch:2006he,Ammon:2010pg,Abt:2019tas,Nakas:2020hyo}.

In this study, we consider a toy model with a fermionic field $\psi$  to utilize the D7's induced metric. 
One should notice that the brane configuration is coming from the D7-brane and, therefore, common to all the  fluctuations of it. 
Our fermionic field is coupled to  the  bulk $\mathrm{U}(1)$   gauge fields   as in the bottom-up models, e.g., \cite{Liu:2009dm,Faulkner:2009wj}. 
%The probe fermionic field $\psi$ with a bulk mass $m$ is related to a fermionic operator $\mathcal{O}_{\psi}$ with a conformal dimension of $\Delta =2+m$ in dual theory.
%Since we focus on the deconfinement phase which corresponds to the black hole embedding in the D3-D7 model, we expect that the probe fermionic operator can be understood as the quark-like operator in the deconfinement phase.
%Note that our fermionic field is different from a fermionic field appearing in the fermionic action of the D7-brane such as those studied in refs.~\cite{Martucci:2005rb,Kirsch:2006he,Ammon:2010pg,Abt:2019tas}.
%That fermionic field correspond to the mesino in the boundary field theory with supersymmetry.
%One of crucial differences between our probe fermionic field and the mesino fluctuation is that the mesino fluctuation has no charge of the worldvolume $\mathrm{U}(1)$ gauge transformation because the mesino is neutral as same as the meson.
%(In \cite{Kirsch:2006he}, the mesino fluctuation is coupled with the isospin gauge field.)
%The other difference is that we just consider only a AdS$_5$-like part of the 8 dimensional induced metric because there are problems with considering full 8 (or 10) dimensional theory. See Appendix \ref{}.
%The normal phase of the high-$T_c$ superconductors will be expected to be strange metal.
%In \cite{Jeong:2018tua}, they found such behavior from the bottom-up model.
For a given embedding, 
 we obtain the fermionic spectral functions by solving the Dirac equation.
Because we consider the system in the black-hole geometry, the Fermi surface is always smeared.
We can locate the smeared Fermi surface  by the pole or singularity position of the spectral function and  the density of state has a Drude-like peak at zero frequency with a finite width.
%We investigate the behavior of the width at various temperature and we find the qualitatively change of the width corresponding to the type of the brane embeddings: the width is strongly suppressed when the embedding belongs to the BH-I phase, and the width becomes relatively large when the embedding belongs to the BH-II phase.
We study the behavior of the decay width of the fermion for various temperatures.
The width exhibits a universal behavior of holographic models at   high enough  temperature.
In a specific range of the parameters,  it has a jump in the temperature  associated with the  first order phase transition of the background D3-D7 system.  We believe that this is universal for all other brane models. 

From the fermionic spectral functions, we study the temperature dependence of the decay rate  and we observe a jump in it at the critical temperature that corresponds to the first order phase transition. We found that the jump in the decay rate mimics that in the resistivity  data in a recent  heavy fermion material.  
If the material were weakly interacting, this would be natural, but it is unclear why this similarity 
holds in the strongly interacting case, and  understanding this  is left as an future work. 
  
This paper is organized as follows.
We present a review of the D3-D7 model with finite density in Sec.~\ref{sec:review_D3D7}.
In Sec.~\ref{sec:probe_fermion}, we consider a toy model of the fermionic field mentioned above probing the  background D3-D7-brane system, and we study the spectral functions of the dual operator.
We also study the width of the spectral functions and its temperature dependence and we compare it with the resistivity data.  We   discuss and conclude   in Sec.~\ref{sec:discussions}.

\section{A review: D3-D7 model with finite density}\label{sec:review_D3D7}
\subsection{Background solutions}
We briefly review the D3-D7 model \cite{Karch:2002sh} with finite baryon density and   temperature following \cite{Nakamura:2006xk,Nakamura:2007nx} where the authors showed phase transitions from a black-hole phase to another black-hole phase. 
The action of the D7 probe brane is given by the following Dirac-Born-Infeld action
\begin{equation}
	S_{\text{DBI}}[X, A] = - \tau_7 \int\dd[8]{\xi}
	\sqrt{-\det[h_{ab} + (2\pi\alpha')F_{ab}]},
\end{equation}
where $\tau_7$ is the tension of the D7-brane and $h_{ab}$ is the induced metric   given by
\begin{equation}
	h_{ab} = \pdv{X^M}{\xi^a}\pdv{X^N}{\xi^b} g_{MN}.
\end{equation}
$\xi^a$ and $X^M$ are the coordinates of worldvolume  and ten-dimensional bulk, respectively.
$g_{MN}$ is the metric of the background ten-dimensional spacetime.
We set the background spacetime to Schwarzschild-AdS$_5\times \mathrm{S}^5$ spacetime.
In isotropic coordinates, the metric can be written as
\begin{equation}
\begin{aligned}
	\dd{s}^2 =&
%	\frac{w^2}{L^2}\left[
%		-\frac{f_1(w)^2}{f_2(w)}\dd{t}^2 + f_2(w)\dd{\vec{x}}^2
%	\right]
%	+
%	\frac{L^2}{w^2}\left(
%		\dd w^2 + w^2 \dd\Omega_5^2
%	\right)\\
%	=&
	\frac{w^2}{L^2}\left[
		-\frac{f_1(w)^2}{f_2(w)}\dd{t}^2 + f_2(w)\dd{\vec{x}}^2
	\right]
	+
	\frac{L^2}{w^2}
	\left(
		\dd{\rho}^2 +  \rho^2 \dd{\Omega_3}^2
		+ \dd{w_5}^2 + \dd{w_6}^2
	\right),
\end{aligned}
\label{eq:metric_isotropic}
\end{equation}
where $f_{1}(w) = 1 - w_h^4/w^{4}$, $f_{2}(w) = 1 + w_h^4/w^{4}$,  $w^2 = \rho^2 + w_5^2 + w_6^2$,
and $\dd{\Omega_3}^2$ is the  line element of the unit three-sphere.%
\footnote{
	The radial coordinate $w$ can be written as $w^2 = \sum_{i=1}^6 {w_i}^2$ and $\rho^2 = \sum_{i=1}^4 {w_i}^2$.
	The coordinate $w$ is related to the Schwarzschild coordinate $r$ by $w^2 = \frac{r^2}{2} + \frac{1}{2}\sqrt{r^4 - r_h^4}$.
}
The Hawking temperature $T$ is related to the location of the horizon $w_h$ by $w_h = \pi T / \sqrt{2}$.
For convenience, we use the following metric
\begin{equation}
	\dd{s}^2 =
	\frac{L^2}{u^2}\left[
		-\frac{f(u)^2}{\tilde{f}(u)}\dd{t}^2 + \tilde{f}(u)\dd{x}^2
	\right]
	+
	L^2
	\frac{\dd{u}^2}{u^2}
	+ L^2 \left(
		\dd{\theta}^2 + \sin^2\theta \dd{\varphi}^2 + \cos^2\theta \dd{\Omega_3}^2
	\right),
\end{equation}
where $f(u) = 1 - u^4/u_h^4$, $\tilde{f}(u) = 1+ u^4/u_h^4$.
$u$ and $w$ are related by $u=1/w$.
$w_5, w_6$, and $\rho$ are related to $\theta,\varphi$, and $u$ by
\begin{equation}
	w_5 = u^{-1} \sin\theta\cos\varphi,\quad
	w_6 = u^{-1} \sin\theta\sin\varphi,\quad
	\rho = u^{-1} \cos \theta,
	\label{eq:relation_of_coordinates}
\end{equation}
respectively.
The other nonzero supergravity field is a Ramond-Ramond five-form flux
\begin{equation}
	F_{(5)} =
	- \frac{4}{L^4 u^5} f(u) \tilde{f}(u) \dd{t}\wedge\dd{x}\wedge\dd{y}\wedge\dd{z}\wedge\dd{w}
	+ 4L^4 \mathrm{vol}(\mathrm{S}^5),
	\label{eq:RR_5-form}
\end{equation}
where $\mathrm{vol}(\mathrm{S}^5)$ is the volume form of the unit five-sphere.
It satisfies the self-dual constraint in the type IIB supergravity: $\ast F_{(5)} = F_{(5)}$.

We choose the worldvolume coordinates as $\xi^a = (t,\vec{x},u, \Omega_3)$, then the transverse directions are $\theta$ and $\varphi$.
Since $\varphi$ can be set to zero by virtue of a symmetry, $\theta(u)$ describes an embedding of the probe brane.
The induced metric is given by
\begin{equation}
	h_{ab}\dd{\xi^a}\dd{\xi^b}
	=
	\frac{1}{u^2} \left[
		-\frac{f(u)^2}{\tilde{f}(u)}\dd{t}^2 + \tilde{f}(u)\dd{x}^2
	\right]
	+
	\left(
		\frac{1}{u^2}
		+ \theta'(u)^2
	\right)\dd{u}^2
	+  \cos^2\theta \dd{\Omega_3}^2.
	\label{eq:induced_metric}
\end{equation}
We set $L=1$ for simplicity.
$F_{ab}$ is the field strength of the worldvolume $\mathrm{U}(1)$ gauge fields $A_a$.
We consider an ansatz for the gauge fields as $A_a \dd{\xi}^a = A_t(u) \dd{t}$ so the nonzero components of the field strength are only $F_{tu}$ and $F_{ut}$.
Writing $S_{\text{DBI}} = \int \mathcal{L}$, the Lagrangian density is given by
\begin{equation}
	\mathcal{L} = - \mathcal{N} \cos^3\theta h_{xx}^{3/2}
	\sqrt{|h_{tt}| h_{uu} - A_t'(u)^2},
	\label{eq:DBI_Lagrangian}
\end{equation}
where $\mathcal{N} = \tau_7 (2\pi^2)$.
A constant of motion $d$ for $A_t$ is given by
\begin{equation}
	d = \frac{1}{\mathcal{N}}\pdv{\mathcal{L}}{A_t'(u)}.
\end{equation}
$d$ is related to the charge density in the boundary theory.
Solving the equation of motion, we can write
\begin{equation}
	A_t'(u) =
	%- d \sqrt{
	%	\frac{|h_{tt}|h_{uu}}{d^2 + \cos^6 \theta^3}
	%}
	%=
	- d u f(u)
	\sqrt{
		\frac{
			1+u^2 \theta'(u)^2
		}{
			\tilde{f}(u)(d^2 u^6 + \tilde{f}(u)^3\cos^6\theta(u))
		}
	}.
\end{equation}
A chemical potential $\mu$ is obtained by
\begin{equation}
	\mu = \int_{u_h}^{0} A_t'(u) \dd{u} = A_t(0) - A_t(u_h).
\end{equation}
We have set $A_t(u_h) = 0$.
%\footnote{
	%In the case of $T=0$ and $\theta(u)=0$, we can perform the integration then we obtain the relation between $\mu$ and $d$ by $\mu = C d^{1/3}$ where
	%$
	%	C = \frac{\Gamma(1/6)\Gamma(4/3)}{2\sqrt{\pi}} \approx 1.4022.
	%$
	%This relation will be valid when $\mu$ or $d$ is enough larger than any other scales.
%}

We perform the Legendre transformation to eliminate $A_t'(u)$ from the Lagrangian density (\ref{eq:DBI_Lagrangian}):
\begin{equation}
	\tilde{\mathcal{L}}
	\equiv
	\mathcal{L} - A_t'(u) \pdv{\mathcal{L}}{A_t'(u)}
	= - \mathcal{N}\sqrt{
		|h_{tt}|h_{uu} (d^2 + h_{xx}^3 \cos^6\theta)
	}.
	\label{eq:Legendre_Lagrangian}
\end{equation}
The Euler-Lagrange equation of $\tilde{\mathcal{L}}$ gives us the equation of motion for $\theta(u)$:
\begin{equation}
	\pdv{u}\left[
		-\sqrt{\frac{|h_{tt}| \Xi }{h_{uu}}} \theta'(u)
	\right]
	- 3 \cos^5\theta\sin\theta h_{xx}^3
	\sqrt{
		\frac{|h_{tt}| h_{uu}}{\Xi}
	}= 0,\quad
	\Xi \equiv d^2 + h_{xx}^3 \cos^6\theta.
	\label{eq:ODE_theta}
\end{equation}
We can solve the equation of motion from $u=u_h$ with a regular condition.%
\footnote{
	In the case of $T=0$, the analytic solutions for the embedding function and the gauge field are found in Ref.~\cite{Karch:2007br}.
}
For regular solutions, $\theta'(u_h) = 0$ must be satisfied at $u=u_h$.
The family of solutions is parametrized by $\theta(u_h)$ and $d$.
The range of $\theta(u_h)$ is $0\leq\theta(u_h)<\pi/2$.
$\theta(u)$ has an asymptotic expansion of
\begin{equation}
	\theta(u) = m_q u + \theta_2 u^2 + \cdots,
	\label{eq:theta_asymptotic}
\end{equation}
at $u=0$.
$m_q$ and $\theta_2$ are related to the quark mass and the quark condensate in the boundary theory, respectively.
[See eq.~(\ref{eq:rel_mass_condensate}).]
In the isotropic coordinates, $m_q$ is the separation distance between the probe D7-brane and the D3-branes at the AdS boundary: $w_5(u=0)=m_q$.

Since the system has a scaling symmetry, we should consider only scale invariants.
By taking $m_q$ as a scale, we define scale invariant temperature, chemical potential and density by
\begin{equation}
	\tilde{T} \equiv \frac{T}{m_q},\quad
	\tilde{\mu} \equiv \frac{\mu}{m_q},\quad
	\tilde{d} \equiv \frac{d}{m_q^3},
\end{equation}
respectively.
We also define scaled isotropic coordinates by $\tilde{w}_5=w_5/m_q$, $\tilde{w}_6 = w_6/m_q$ and $\tilde{\rho} = \rho/m_q$ for fixed $m_q$.
By definition, $\tilde{w}_5(u=0)$ is always one.

\begin{figure}[htbp]
	\centering
	\includegraphics[width=0.82\linewidth]{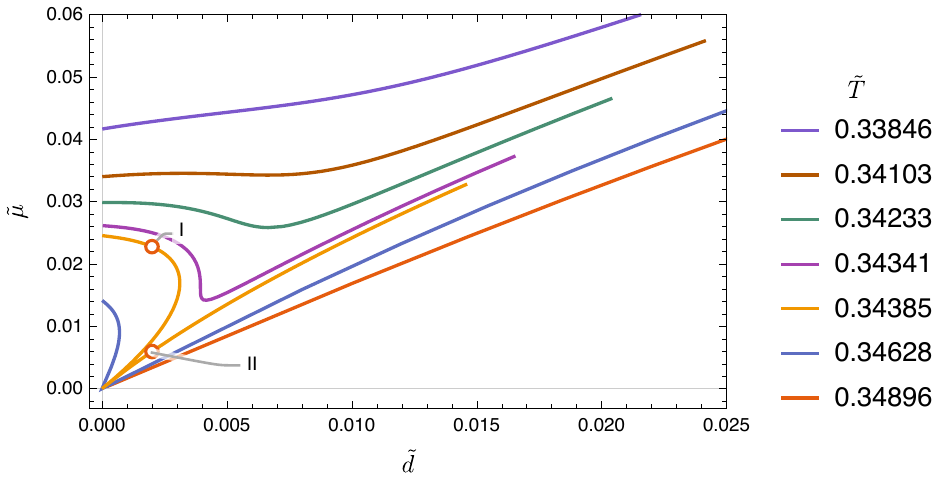}
	\includegraphics[width=0.56\linewidth]{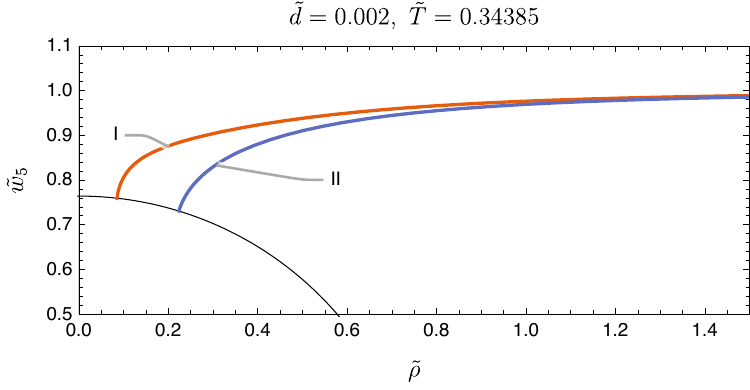}
	\caption{
		Top: relation between the density and chemical potential for various temperatures.
		Bottom: brane embeddings corresponding to I and II in the top panel.
	}
	\label{fig:d-mu}
\end{figure}
By solving Eq.~(\ref{eq:ODE_theta}), we obtain relation between $\tilde{d}$ and $\tilde{\mu}$ for several $\tilde{T}$, as we show in the top panel of Fig.~\ref{fig:d-mu}.
The results agree with those obtained in Ref.~\cite{Nakamura:2007nx}.
In the range of $0.34341<\tilde{T}<0.34468$, $\tilde{\mu}$ becomes a multivalued function of $\tilde{d}$, and, hence, there are multiple solutions for given $\tilde{d}$ at $\tilde{T}$.
In the bottom panel of Fig.~\ref{fig:d-mu}, we show the corresponding embeddings labeled by I and II at $\tilde{d}=0.002$ and $\tilde{T} = 0.34385$.%
\footnote{
	Since the solution at the middle point of the $\tilde{\mu}$--$\tilde{d}$ curve will be unstable, we do not focus on it in this paper.
}
We refer the upper solution and the lower solution in the bottom panel of figure \ref{fig:d-mu} as  BH-I and II embedding, respectively.
The first order phase transition occurs in such a case.
The transition points are determined from the free energy or the Maxwell construction as we discuss in Appendix \ref{appendix:free_energy}.
%We call this transition as a geometrical phase transition.
At $\tilde{T} = 0.34341$, the system has the second order phase transition when $\tilde{d} = 0.0039385$.
At a temperature out of the above range, there is a crossover.

\begin{figure}[htbp]
	\centering
	\includegraphics[width=0.49\linewidth]{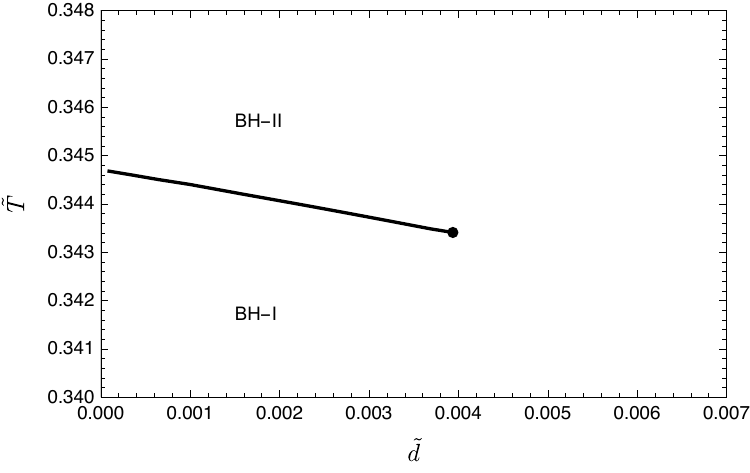}
	\includegraphics[width=0.49\linewidth]{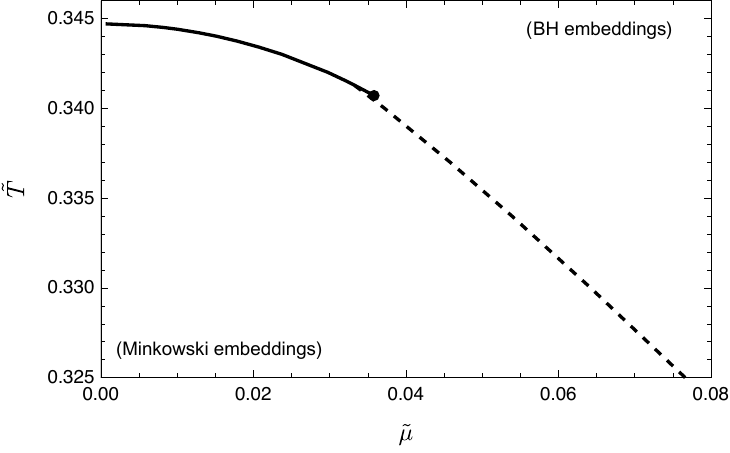}
	\caption{
		Phase diagrams of the probe brane.
		Left: canonical ensemble, i.e., $\tilde{d}$ is a controlling parameter.
		The black line is the first order phase transition line.
		The end point at $\tilde{d} = \tilde{d}_c \approx 0.004$ indicates the second order phase transition end point.
		For $\tilde{d}<\tilde{d}_c$, we call the low- and high-temperature regime as the BH-I and BH-II phase, respectively.
		Right: grand canonical ensemble; i.e., $\tilde{\mu}$ is a controlling parameter.
		The black line is the first order phase transition line.
		The dashed line denotes the second order phase transition line.
	}
	\label{fig:geo_phase_diagram}
\end{figure}

The phase structures are summarized as phase diagrams in Fig.~\ref{fig:geo_phase_diagram}.
The phase structure changes depending on whether $\tilde{d}$ or $\tilde{\mu}$ is treated as a controlling parameter, in other words, considering a canonical ensemble or a grand canonical ensemble, respectively.
We show the first order transition line as the black curve, and the second order phase transition line as the black dashed line.
In the canonical ensemble, the ranges of the phase transition are $0.34341<\tilde{T}<0.34468$ and $0<\tilde{d}<\tilde{d}_c=0.0039385$.
In this case, the embeddings are always given by the BH embeddings but it is divided by the first order phase transition line in $\tilde{T}$ for $\tilde{d}<\tilde{d}_c$.
We call the low- and high-temperature regime in $\tilde{d}<\tilde{d}_c$ as the BH-I and BH-II phase, respectively.
The BH-I and II embeddings shown in the bottom panel of Fig.~\ref{fig:d-mu} belong to the BH-I and II phase, respectively.
In the grand canonical ensemble, there are brane solutions without touching the black-hole horizon called Minkowski embeddings.
The Minkowski embeddings are realized with vanishing density.
We do not focus on such solutions in this study.

\section{Probing by a fermionic field}\label{sec:probe_fermion}
{%\color{blue}
In this section, we consider dynamics of a spinor field probing the background D7-brane's worldvolume. 
 Our fermion is   of bottom up character  introduced in the contexty of  D3-D7.
In the scaling limit where D3's  disappear  to be the AdS gravity,  the string connects D7 and horizon, and the low-energy physics of the string  becomes that of  the Dirac fermion in  fundamental  $U(N_f)$ representation on D7 world volume. Here, we set $N_f=1$. This string as a fermion was also utilized in 
\cite{Mateos:2007vc} to argue that, in the presence of the finite density of the fermion, only black-hole embedding (BHE) is allowed.  Here, our transition is from BHE to BHE. 
The  Dirac fermion in fundamental representation is charged and, therefore,  should contribute to the conductivity.
As a first step of the study of the fermionic spectral function in the D3-D7 model, we consider the fermion's action governed  by the five-dimensional part of the induced metric (\ref{eq:induced_metric}) ignoring the  three-sphere part of induced  metric to avoid the technical complication. This should not change the essential features, since
the shape of the brane is already encoded in the five-dimensional model and extra factors of $\cos\theta$ in the measure should not change the qualitative features of the theory. 
%We also present discussion of the other formulation
%and the problem arising there
%in Appendix \ref{apdx:fermion_in_worldvolume}.
}

We now consider the following simplified model:
\begin{equation}
	S_{\text{spinor}} = i \int\dd[5]{x}\sqrt{-\det h_{\mu\nu}} \bar{\psi} \left(
		\gamma^{\mu} \mathcal{D}_{\mu} - m
	\right) \psi + S_{\text{bdy}},
\end{equation}
where $h_{\mu\nu}$ is the five-dimensional part of the induced metric, that is,
\begin{equation}
	h_{\mu\nu}\dd{x^{\mu}}\dd{x^{\nu}}
	=
	\frac{1}{u^2} \left[
		-\frac{f(u)^2}{\tilde{f}(u)}\dd{t}^2 + \tilde{f}(u)\dd{x}^2
	\right]
	+
	\left(
		\frac{1}{u^2}
		+ \theta'(u)^2
	\right)\dd{u}^2.
\end{equation}
$\mathcal{D}_{\mu} = \nabla_{\mu} - i q A_{\mu}$ is a gauge covariant derivative, and $S_{\text{bdy}}$ is the boundary action  which will be specified later. 
Notice that the covariant derivative is also defined with respect to the five-dimensional metric $h_{\mu\nu}$.
Using the spin connection with respect to $h_{\mu\nu}$, it can be written as
\begin{equation}
	\nabla_{\mu} = \partial_{\mu}
	+ \frac{1}{8} \omega_{\mu}{}^{\underline{\nu\rho}}
	[\gamma_{\underline{\nu}}, \gamma_{\underline{\rho}}], 
\end{equation}
where $\gamma^{\mu}$ denotes gamma matrices in the curved spacetime and  $\gamma_{\underline{\mu}}$ are gamma matrices in the tangent space that will be defined as follows: It can be written as
$
	\gamma^{\mu} = e^{\mu}{}_{\underline{\mu}} \gamma^{\underline{\mu}},
$
where $e^{\mu}{}_{\underline{\mu}}$ is the inverse matrix of vielbein $e_{\mu}{}^{\underline{\mu}}$ which satisfies $e_{\mu}{}^{\underline{\mu}} e_{\nu}{}^{\underline{\nu}} \eta_{\underline{\mu\nu}} = h_{\mu\nu}$, and $\gamma^{\underline{\mu}}$ is the gamma matrices in the tangent space.
The gamma matrices in the five-dimensional spacetime can be chosen as
\begin{equation}
	\gamma^{\underline{0}} = \sigma^1 \otimes i \sigma^2,\quad
	\gamma^{\underline{1}} = \sigma^1 \otimes \sigma^1,\quad
	\gamma^{\underline{2}} = \sigma^1 \otimes \sigma^3,\quad
	\gamma^{\underline{3}} = \sigma^2 \otimes I_2,\quad
	\gamma^{\underline{u}} = \sigma^3 \otimes I_2,
	\label{eq:5d_gamma_matrices}
\end{equation}
where $\sigma^{1},\sigma^{2},$ and $\sigma^{3}$ are the Pauli matrices.
Then, $\psi(x^{\mu})$ is written as a four-components spinor field.

The equation of motion is  the   Dirac equation
\begin{equation}
	\left(\gamma^{\mu}\mathcal{D}_{\mu} - m\right) \psi(t, \vec{x}, u) =0.
\end{equation}
Substituting the five-dimensional metric, we can write
\begin{equation}
	\gamma^{\mu}\mathcal{D}_{\mu} =
	{e^{\nu}}_{\underline{\mu}} \gamma^{\underline{\mu}} (\partial_{\mu} - i q A_{\mu})
	+ \frac{1}{4} {e^u}_{\underline{u}} \gamma^{\underline{u}}
	\partial_u \ln(- \det(h_{\mu\nu}) h^{uu}).
\end{equation}
Considering the following transformation:
\begin{equation}
	\psi(x^{\mu}) = \left(-\det(h_{\mu\nu}) h^{uu} \right)^{-1/4} \phi(x^{\mu}),
\end{equation}
we obtain the following equation:
\begin{equation}
	\left[
		{e^{\nu}}_{\underline{\mu}} \gamma^{\bar{\mu}} (\partial_{\nu} - i q A_{\nu}) - m
	\right]\phi(x^{\mu}) = 0.
\end{equation}

We decompose the four-component spinor field $\psi$ as follows:
\begin{equation}
	\psi(x^{\mu})=
	\psi_{+}(t, \vec{x} ,u ) +
	\psi_{-}(t, \vec{x} ,u ),
\end{equation}
where $\psi_{\pm}$ are projected by $\psi_{\pm} = P_{\pm}\psi$ with $P_{\pm} = (1\pm \gamma^{u})/2$.
%The each terms are eigenvector of $\gamma^{u}$ with eigenvalues of $\pm1$.
According to Ref.~\cite{Iqbal:2009fd}, the asymptotic behaviors of the spinors are written as
\begin{equation}
	\psi_{+} = \psi_{+}^{(0)} u^{\Delta_{-}} + \psi_{+}^{(1)} u^{1 + \Delta_{+}} + \cdots,\quad
	\psi_{-} =  \psi_{-}^{(0)} u^{\Delta_{+}} + \psi_{-}^{(1)} u^{1 + \Delta_{-}}\cdots,
\end{equation}
where $\Delta_{\pm} = 2 \pm m$.
%\footnote{
%	Note that the redefined field is related to th original field by $\psi \approx u^{-2} \phi$ in the vicinity of $u=0$.
%}
$\psi_{\pm}^{(0)}$ and $\psi_{\pm}^{(1)}$ are related to the source and fermionic operator with the scaling dimension of  $\Delta_{+}$, respectively.
To obtain retarded responses, we also impose the ingoing-wave boundary condition at the black-hole horizon.
In order to compute the Green's function, we need to fix the boundary action. We employ
\begin{equation}
	S_{\text{bdy}} = \lim_{u\to \varepsilon} \frac{i}{2} \int \dd[4]x \sqrt{- h h^{uu}} \bar{\psi} \psi,
\end{equation}
where $\varepsilon$ is a small positive cutoff $\varepsilon$, and $\bar{\psi}_{+} = \psi^{\dagger} \gamma^{0}$.%
\footnote{
	A similar boundary action was employed in \cite{Ammon:2010pg} for the top-down model.
}
This choice of the boundary term is known as the standard quantization \cite{Laia:2011zn}.
The retarded Green's function is obtained by
\begin{equation}
	G^{\text{R}}(k) = i \mathcal{S} \gamma^{0} = \frac{2 m + 1}{k^2} (\gamma \cdot k) \mathcal{T} \gamma^{0},
\end{equation}
where $\mathcal{S}$ and $\mathcal{T}$ are defined, respectively, by
\begin{equation}
	\psi_{-}^{(1)} = \mathcal{S} \psi_{+}^{(0)},\quad
	\psi_{+}^{(1)} = \mathcal{T} \psi_{-}^{(0)}.
\end{equation}
The Green's function has a scaling dimension of $-2m$.
We can also derive the flow equation for $G^{\mathrm{R}}(k, u)$ from the Dirac equation, as those in Ref.~\cite{Yuk:2022lof}.
In the following section, we compute the result by solving the flow equation.

For later convenience, we define spectral function by 
\begin{equation}
	A(\omega, |\vec{k}|) = - \mathrm{Im} \tr G^{\mathrm{R}}(\omega, \vec{k}).
\end{equation}
Since the system is isotropic, $A(\omega, k)$ depends only on $k \equiv |\vec{k}|$.
We also define a scaled spectral function
\begin{equation}
	\tilde{A}(\tilde{\omega},\tilde{k}) = m_q^{2 m} \times A(m_q \tilde{\omega}, m_q \tilde{k}).
\end{equation}

\subsection{Spectral function}
\begin{figure}[htbp]
	\centering
	\includegraphics[width=0.45\linewidth]{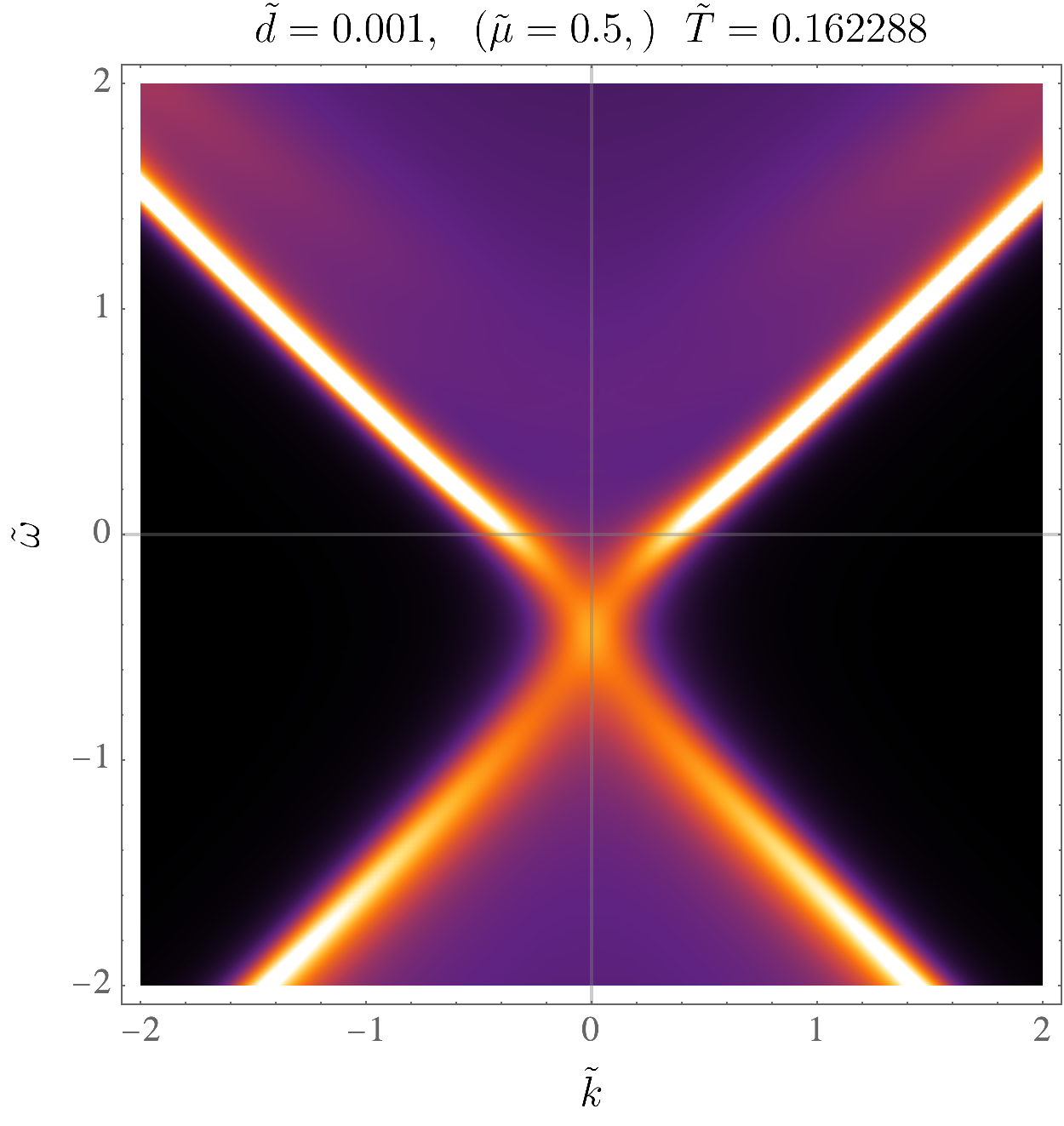}
	\includegraphics[width=0.45\linewidth]{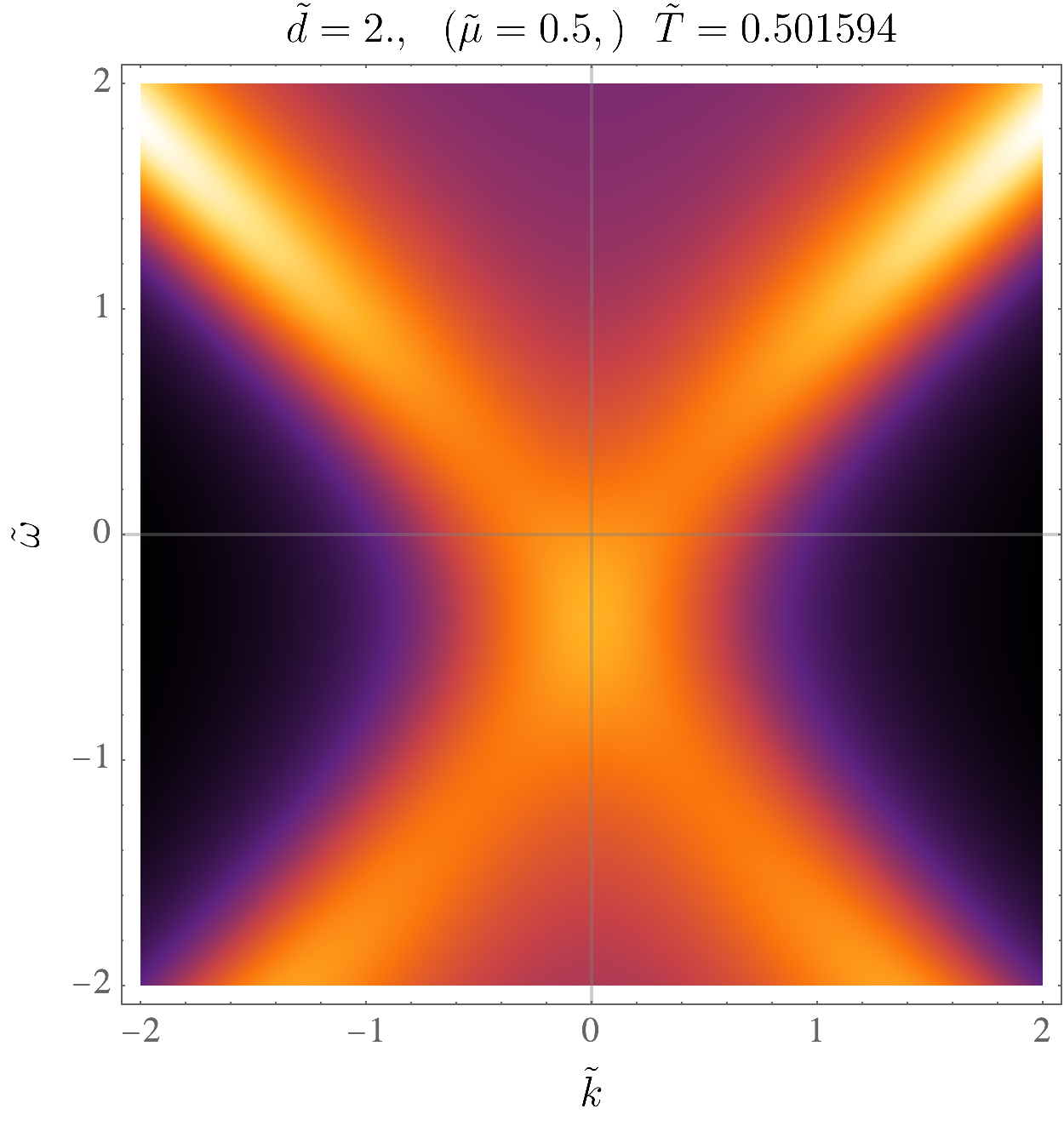}\\
	\includegraphics[width=0.45\linewidth]{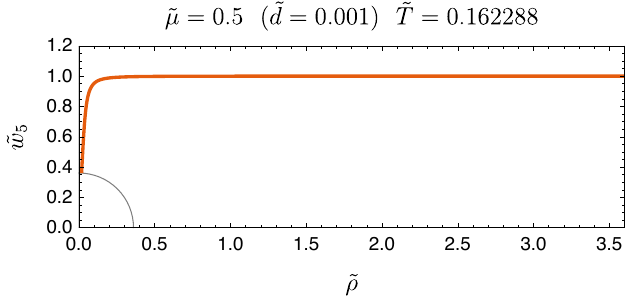}
	\includegraphics[width=0.45\linewidth]{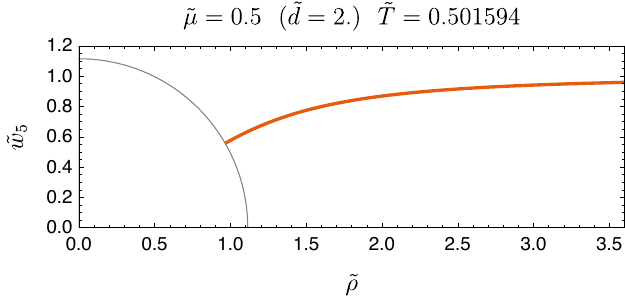}
	\caption{
		Top: spectral functions $\tilde{A}(\tilde{\omega},\tilde{k})$ for two setups.
		We set $m=0.2$.
		Bottom: brane embeddings corresponding to the top panels, respectively. The black circle shows the location of the black-hole horizon.
	}
	\label{fig:spectral_func}
\end{figure}
We show the spectral functions of the two embeddings with $\tilde{\mu}=0.5$ in Fig.~\ref{fig:spectral_func}.
The left panel in Fig.~\ref{fig:spectral_func} corresponds to a solution in the BH-I phase, and the right panel in Fig.~\ref{fig:spectral_func} corresponds to a solution in the crossover region in Fig.~\ref{fig:geo_phase_diagram}.
The spectral function of the left panel in Fig.~\ref{fig:spectral_func} is similar to those obtained in Ref.~\cite{Cubrovic:2009ye}.
In both cases, the Dirac points are shifted by $\mu$.
It is also similar to the results of Refs.~\cite{Liu:2009dm,Cubrovic:2009ye}.
As $T$ increases, the peaks of the spectral function are smeared.
%The higher $T$ increases, the broader the width of the peaks in the spectral function.

It is considered that the Fermi level is located at $\omega = 0$.
%\footnote{
%	The Fermi level is measured via the Fermi-Dirac distribution
%	\(
%		f_F(E) = \frac{1}{e^{(E - E_F)/T} + 1}.
%	\)
%	The word of  ``Fermi energy'' is often used for non-interacting and zero temperature case.
%}
The intersection between the peaks and the horizontal axis is considered as the Fermi surface, but it is smeared at finite temperatures.
In the following, we define the Fermi momentum of the smeared Fermi surface and the width of the Drude-like peak.

\begin{comment}
In our choice of the gammma matrices, $G_{ij}$ is diagonalized when the spatial momentum is along to $x$-direction.
So, we also consider
\begin{equation}
	A_{11}(\omega, k_x) = - \mathrm{Im} G_{11}^{\mathrm{R}}(\omega, k_x, 0, 0).
\end{equation}
$A_{22}$ is also defined by same way but it is just equal to $A_{11}(\omega, -k_x)$ in our setup.
Figure \ref{fig:spectral_func_G11} shows $A_{11}$ for the embedding A'.
We computed the first stable pole location of $A_{11}$ in figure \ref{fig:pole_Aprime} and \ref{fig:pole_B} for the embedding A' and B, respectively.
$\omega_R(k_x)$ is understood as the dispersion relation of the quasiparticle.

The local density of state $D$ should be related to the Green's function by 
\begin{equation}
	D(\omega) = - \frac{1}{\pi} \int\dd[3]{k} \mathrm{Im} \tr G^{\mathrm{R}}(\omega, \vec{k}).
\end{equation}
In our case, however, the integrating over $\vec{k}$ will not be converge.
We consider the following as an analogue of  the density of state.
\begin{equation}
	D'(\omega) \equiv - \frac{1}{\pi} \mathrm{Im} \tr G^{\mathrm{R}}(\omega, |\vec{k}| = k'_{F}).
\end{equation}

By assuming single pole structure of the Green's function, we define the ``self-energy'' $\Sigma$ by
\begin{equation}
	\tr G^{\mathrm{R}}(\omega, \vec{k})
	\equiv \frac{1}{\omega - k - \Sigma(\omega, \vec{k})}.
\end{equation}
\end{comment}

\subsection{Smeared Fermi surface}
At finite temperatures the Fermi surface is smeared, so we can no longer define sharp Fermi momentum $k_{F}$. However, we can still define an analog of $k_{F}$ from the "pole" of the retarded Green's function there.%
\footnote{
	Another possible definition of the smeared Fermi momentum is using local maximum of the spectral function.
	It will be expressed by
	\[
		\{k_F'\} = \arg_k \max A(\omega = 0, k).
	\]
	It determines only $k_F'$. The width will be measured from the peak.
	However, we do not use this definition of the smeared Fermi surface in this study.
}
In normal isotropic metals, $k_F$ satisfies $E(k_F) = \mu$ where $E(k)$ denotes the dispersion relation.
The spectral function in the noninteracting theory should have a delta function peak at the Fermi momentum,
and  the Green's function has a pole  there.  
%\begin{equation}
%	A(\omega = 0, \vec{k}) \propto \delta(k - k_{F}).
%\end{equation}
%We have assumed that $\omega = 0$ is the Fermi level.
%It implies that the Green's function has a pole structure of
%\begin{equation}
%	\mathrm{tr} G^{\mathrm{R}} (\omega, k) = \frac{Z}{\omega - k + k_{F} + i 0_{+}} + \cdots,
%\end{equation}
%where $Z$ is a constant, and the ellipsis denotes contributions from the other poles and the regular part.

At finite temperatures of our interacting theory,  the spectral function is smeared and 
we assume that the pole is located on the lower half complex $\omega$ plane with a finite distance from the real axis so that  the Green's function has the following structure: 
\begin{equation}
	\tr G^{\mathrm{R}}(\omega, k) = \frac{Z_{\Gamma}}{\omega - k  + k_{F}' + i \Gamma/2} + \cdots,
\end{equation}
where $\Gamma$ is  the width, i.e., the decay constant.
$Z_{\Gamma}$ is a constant residue.
$k_{F}'$ is also a positive real constant which can be understood as a center of smeared Fermi momentum.
Taking the inverse of $\tr G^{\mathrm{R}}$ and evaluating it at $(\omega, k_F) = (-i\Gamma/2, k_F')$, we obtain the equation
\begin{equation}
	\frac{1}{\tr G^{\mathrm{R}}(\omega=- i\Gamma/2, k=k_F')}= 0.
	\label{eq:k_F_definition_pole}
\end{equation}
We can determine $k_F'$ and $\Gamma$ by solving this complex-valued equation.
%Actually, we can easiely find a set of $k_F'$ and $\Gamma$ by solving eq.~(\ref{eq:k_F_definition_pole}) numerically.
In the following, we will study the behavior of $\Gamma$ by using Eq.~(\ref{eq:k_F_definition_pole}).

\subsection{$T$ dependence of the decay rate}
\begin{figure}[htbp]
	\centering
	\subfigure[Decay rate of fermion in D3-D7]
	{\includegraphics[width=0.45\linewidth]{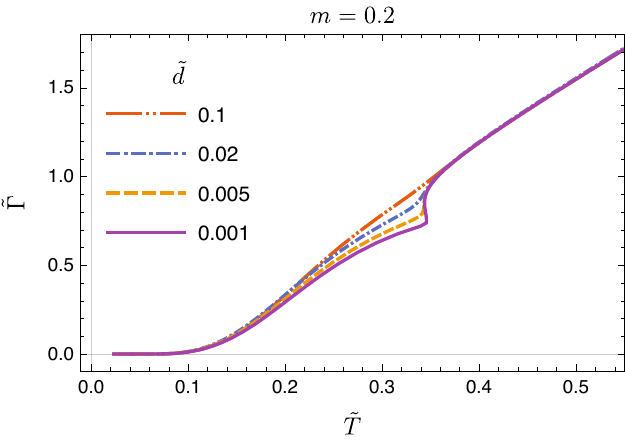}}
	\subfigure[ resistivity  data]
	{\includegraphics[width=0.4\linewidth]{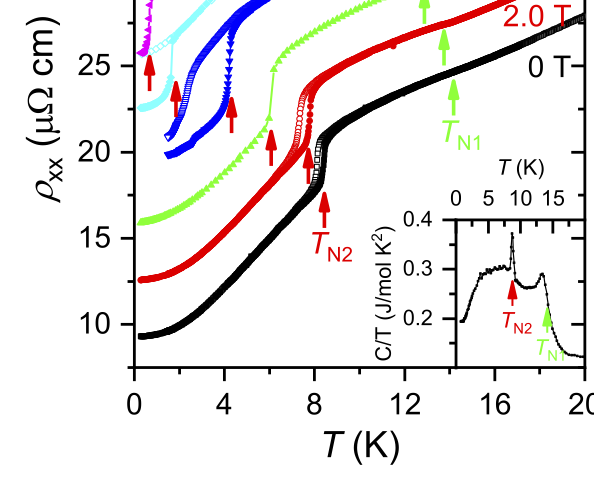}}
		\subfigure[%
	  Hysteresis on   $\tilde{\Gamma}$ %vs temperature $\tilde{T}$ for $\tilde{d}=0.001$.
	]{
		\includegraphics[width=0.45\linewidth]{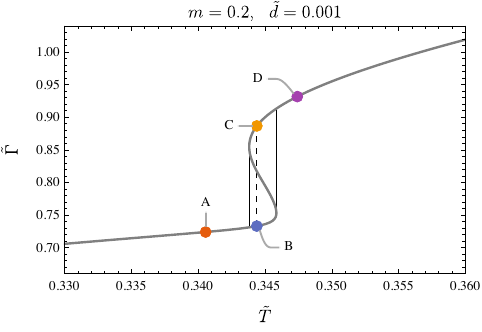}
	}
	\subfigure[Probe-brane's embeddings]{
		\includegraphics[width=0.45\linewidth]{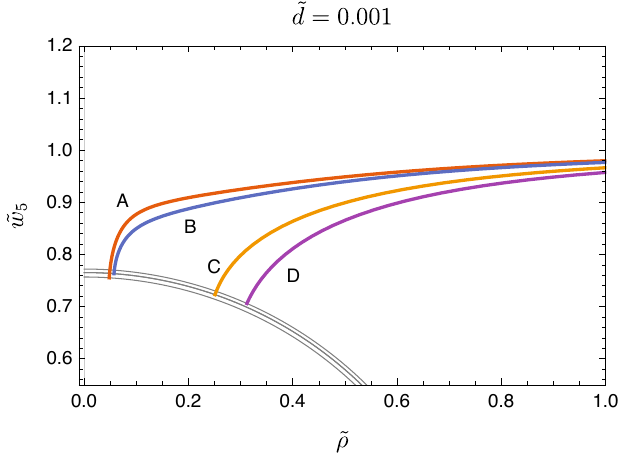}
	}
		\caption{
		(a)$ {\Gamma}$ vs $ {T}$ for various {densities}.  (b) 
		Resistivity data in a heavy fermion material. 
		Curiously, both a) and b) show the  jump from a linear $T$ to another linear in $T$. 
		%The black line is the experimental data with vanishing magnetic fields $B=0$ to be compared with figure(a).
		In (c), the vertical dashed line denotes the  $T_c$. 
	%	The two vertical lines indicate the edges of the two metastable branches corresponding to the hysteresis behavior.
		The points labeled by A -- D correspond to the embeddings in the right panel, respectively.
		In (d), the gray circles denote the black-hole horizon at each temperature.
		The embeddings B and C have a common temperature $\tilde{T} = \tilde{T}_c$.
		(b) is from \cite{Wang:2022fim}. 
	}
	\label{fig:Gamma-T}
\end{figure}
%\begin{figure}[htbp]
%	\centering
%	\subfigure[%
%	Width $\tilde{\Gamma}$ vs temperature $\tilde{T}$ for $\tilde{d}=0.001$.
%	]{
%		\includegraphics[width=6cm]{Gamma-T_zoom.pdf}
%	}
%	\subfigure[Probe-brane's embeddings]{
%		\includegraphics[width=6cm]{embeddings_ABCD.pdf}
%	}
%	\caption{
%		In (a), the vertical dashed line denotes the critical temperature where the first order phase transition occurs.
%		The points labeled by A to D correspond to the embeddings in the right panel, respectively.
%		In (b), the gray circle denotes the black hole horizon at each temperatures.
%		The embeddings B and C have a common temperature $\tilde{T} = \tilde{T}_c$.
%		The figure is from \cite{Maurya_2016,Wang:2022fim}. 
%	}
%	\label{fig:Gamma-T_wt_embeddings}
%\end{figure}
From Eq.~(\ref{eq:k_F_definition_pole}), we compute $\Gamma$ for various temperatures.
We define the scaled width by $\tilde{\Gamma} \equiv \Gamma/m_q$.
Figure \ref{fig:Gamma-T} shows the width ${\Gamma}$ as functions of  the temperature ${T}$ for various  values of scaled density and chemical potential, $\tilde{d}$ and $\tilde{\mu}$.
In both cases, we find that ${\Gamma}$  is linear in $ {T}$ at high temperatures.
For sufficiently low density, the curves have small multivalued region around $ {T} = 0.35m_q$ corresponding to the multivalued results shown in Fig.~\ref{fig:d-mu}(a). 
%We will see the multivalued region for fixed $\tilde{d}$ case further later.
%When $\tilde{\mu}$ is taken to sufficiently small, the background solutions are given by the Minkowski embeddings at low temperatures.
%Then, $\tilde{\Gamma}$ becomes exactly zero. (We did not show these results in the right panel of figure \ref{fig:Gamma-T}.)
%When $\tilde{\mu}$ is taken to sufficiently large value, the the background solutions are given by BH embeddings and $\tilde{\Gamma}$ becomes finite.
%In such cases, 
When $ { T}$ is taken to be sufficiently small, $ {\Gamma}$ depends on $ {T}$ as $ {\Gamma}\approx \gamma e^{-\alpha/ {T}}$ with a positive constant $\alpha$ near zero temperature.
%It implies that the the power dependence of $\tilde{\Gamma}$ on $\tilde{T}$ diverges near zero temperature.
These behaviors along the temperature  may  appear in various holographic models; e.g., see Ref.~\cite{Oh:2021xbe}.

Figure \ref{fig:Gamma-T}(c) shows an enlarged view of Fig.~\ref{fig:Gamma-T}(a) around the phase transition point for $\tilde{d}=0.001$.
The dotted vertical dashed line shows the transition point at $T=T_c = 0.34440$.
%In an adiabatic measurement, the result will have a jump between the point B and C, and the middle S-shaped branch will be skipped.
In adiabatic measurements, it is anticipated that the results will exhibit a discontinuity between points B and C, with the intermediate S-shaped branch being omitted.
We show the brane embeddings for the points labeled by A--D in Fig.~\ref{fig:Gamma-T}(d).
The points A and B belong to the BH-I phase, and C and D belong to the BH-II phase.
%In the regime of $T<T_c$, the width $\tilde{\Gamma}$ is strongly suppressed for the small fixed $\tilde{d}$.
%In contrast to $T<T_c$, $\tilde{\Gamma}$ becomes large in $T>T_c$.
%The change of the behavior of $\tilde{\Gamma}$ corresponds to the phase of the brane embedding.

{%\color{red}
In passing, we  point out that our result in decay rate  showing the first order transition from a black-hole embedding phase to another black-hole embedding  phase exhibits a qualitative similarity to the experimental measurement of the resistivity in a Kondo compound \cite{Maurya_2016,Wang:2022fim}, some of which is captured in 
Fig.~\ref{fig:Gamma-T}(b) showing the experimental result  of  Ref.~\cite{Wang:2022fim}.  The Kondo compound also exhibit a drop of the first order phase transition in the resistivity, and the data certainly suggest that there is a first order phase transition from a strange metal to another strange metal with different slope, although at the present time the microscopic reason for such a transition is not known. 
However, we need to make a caution. 
In weakly interacting cases, comparing the resistivity and the decay rate is justified by the Drude theory, but here there is no established reason to do so. Therefore, understanding the similarity is left as a future work of the community.  The presence of the first order transition is common in the brane system and material system, 
and there is a further similarities list below.
\begin{enumerate}
	\item 
 The first is  the  specific heat whose data are in the inset of Fig.~\ref{fig:Gamma-T}(b). One sees that there is a sharp peak at the position of the phase transition.
It is well consistent  with our calculation, because from Figs.~\ref{fig:free_energy}(left)  and \ref{fig:grand_potential}(left),  the free energy has  slope difference around the phase transition point. 
Since the specific heat is second derivative the free energy, $C_V=-T\frac{\partial^2 F}{\partial T^2}$, our free energy calculation shown in Appendix \ref{appendix:free_energy} implies the delta function peak in $C_V$ vs $T$. 
Notice that the figure is for $F/T^4$ vs $d/T^3$ with fixed $T/m_{q}$, so that the same first order  nature appears for fixed density with varying $T$. 
Such a peak is the main feature of the  data in the figure inset in Fig.~\ref{fig:Gamma-T}(b).   
\item  
We  also find that the presence of the  hysteresis  at zero magnetic field is also consistent with our calculation presented in Fig.~\ref{fig:Gamma-T}(c).  
%\item 
%One might ask whether our theory does not encoded the anti-ferromagnetism. 
%We can argue that the AFM is irrelevant because all the phase transition is within the same  AFM phase so that 
%magnetism has nothing to do with the phase transition dynamics. 
\end{enumerate}
These   indicate that the similarity of the brane embedding and the dynamics of the heavy fermion is something we might  utilize in the future study.  But  we should not consider   our present theory  as a serious  explanation of the phenomena at all.  It is not the purpose of this  paper but an observation of similarity. 

\begin{figure}[htbp]
	\centering
	\includegraphics[width=0.45\linewidth]{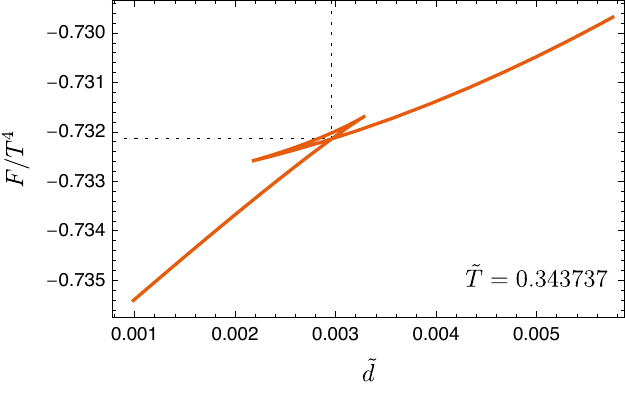}
	\includegraphics[width=0.45\linewidth]{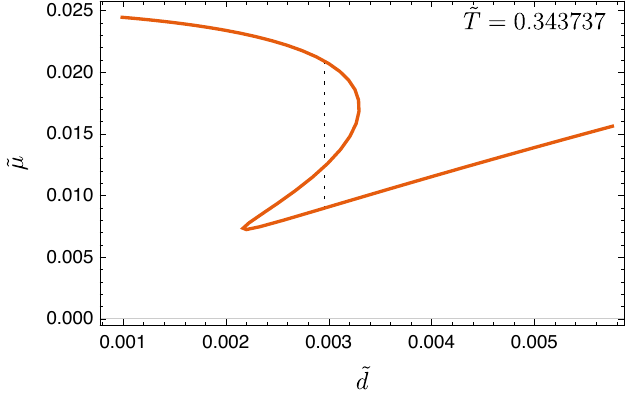}
	\caption{
		Left: free energy vs density at $\tilde{T}=0.343737$.
		The dotted lines shows the first order phase transition point at $\tilde{d}\approx 0.003$.
		Right: $\tilde{\mu}$ vs $\tilde{d}$. The vertical dotted line shows the phase transition point.
	}
	\label{fig:free_energy}
\end{figure}
\begin{figure}[htbp]
	\centering
	\includegraphics[width=0.45\linewidth]{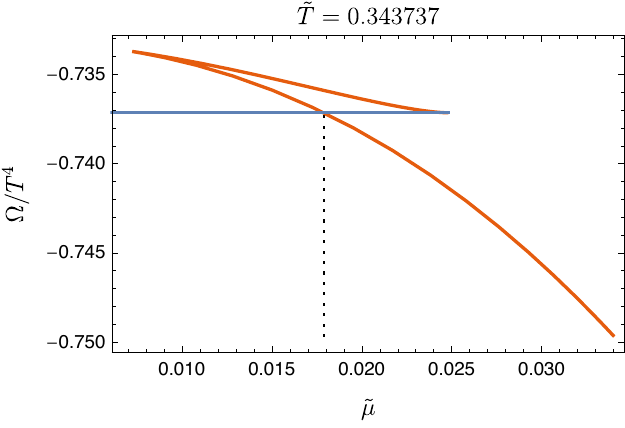}
	\includegraphics[width=0.45\linewidth]{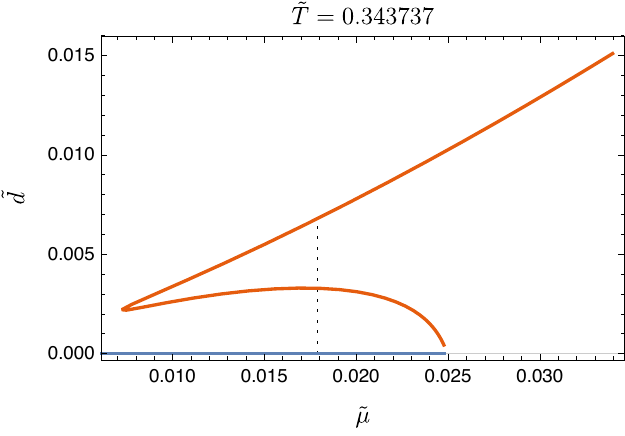}
	\caption{
		Left: grand potential vs chemical potential at $\tilde{T}=0.343737$.
		The dotted line shows the first order phase transition point at $\tilde{\mu}\approx 0.018$.
		The red and blue curves show results in the BH and the Minkowski embedding, respectively.
		Right: $\tilde{d}$ vs $\tilde{\mu}$. The vertical dotted line shows the phase transition point.
	}
	\label{fig:grand_potential}
\end{figure}

We have to make a few  remarks:  
First, our model of the probe fermion is a simplified  model where the extra dimension of the three-sphere is neglected. 
%One may want to analyze the problem by including its role in the future. 
%The ambiguity of the formulation still remains.
%In particular, we dropped the contribution of the 3 sphere.
The treatment of the extra dimensions in the brane should be improved in the future if one truly wants a top-down  theory. 
}

\section{Discussions}\label{sec:discussions}
In this paper, we study the fermionic spectral function in the D3-D7 model by considering the toy model of the probe spinor field. 
From the spectral function, we investigate the behavior of the decay rate for varying temperature.
Most of the quantity we calculated shows a remnant of the first order phase transition. 
We find that the  decay rate  also shows the dropping behavior corresponding to the first order phase transition between the BH-I and BH-II phases.    We also mentioned the similarity of the decay rate to the transport data, although its ground is unclear to us from the strongly coupled system point of view. 
 
It may be  partially related to the puzzle in the holography: While the transport coefficients calculated with holographic method are too sensitive to the details of the background, those in the metallic phase of real condensed matter with strong correlations are universal which exhibit the linear in $T$ resistivity.
It may be useful to remind that the fermion width is shown to be universal.\cite{Yuk:2022lof}
%At this moment, it is not clear what is missing point in the general method of the holographic calculation of the transport.
%Therefore, in the meantime, we should find a bypass track like the method we adopted: use the fermion width with the Drude picture.
%
%We find that there is a qualitative change in the behavior of the width corresponding to the two types of the BH embeddings named by BH-I and BH-II embeddings.

%We found that if we assume that the Drude model works in this case so that the resistivity is proportional to the fermion decay rate,   the jump matches the resistivity  data in a heavy fermion material.  

{%\color{red}
The first order phase transition is one of the characteristic behaviors in the brane models.
 While  the holographic superconductors exhibit only the second order phase transition, the brane models often show the first order phase transition.
 Our original motivation was to understand the physical meaning of the  first order phase transition between  the two black-hole phases   using the probe fermion.
%Indeed, in the measurement of the Kondo compounds \cite{Maurya_2016,Wang:2022fim}, such a feature was observed in the temperature dependence of the resistivity as shown in section \ref{sec:probe_fermion}.  We found the similar behavior in the width of the fermionic spectral function in our model. 

We expect that there is a common mechanism existing for the first order transition between two dissipative phases in both the D3-D7 model and those Kondo compounds.
As far as we know, the physical meaning of the low-temperature phase and the first order phase transition is still not understood in the material science point of view.
It would be very interesting if we can reveal the above point by further investigation.
}

\section*{Acknowledgments}
  We thank the APCTP for the hospitality during the focus program, where part of this work was discussed.
This work is partly supported by NSFC, China (No.~12275166 and No.~12147158).
%S.\,I.\, is supported by National Natural Science Foundation of China with Grant No.~12147158.
S.-J.S. and T.Y. are supported by 
National Research Foundation of Korea Grant No.~NRF-2021R1A2B5B02002603,  No.~NRF-2022H1D3A3A01077468, and No.~RS-2023-00218998 of the Basic research Laboratory support program.

\appendix

\section{The free energy and phase transition points}\label{appendix:free_energy}
In this section, we discuss the free energy of the probe brane and the phase transitions.
The phase transition points can be determined from the thermodynamics of the probe brane.
We interpret the on-shell action of Eq.\ (\ref{eq:Legendre_Lagrangian}) as a Helmholtz free energy \cite{Nakamura:2007nx}:
\begin{equation}
	F_0(d) = \int_{u_h}^{\epsilon} \tilde{\mathcal{L}}(d)\dd{u}
\end{equation}
Since this integral is still divergent, we have to regularize it.
According to \cite{Karch:2005ms} which is equivalent to the procedure in \cite{Nakamura:2006xk,Nakamura:2007nx}, the counterterms are given by
\begin{equation}
	L_1 = \frac{1}{4}\mathcal{N}\sqrt{-\gamma},\quad
	L_2 = -\frac{1}{2}\mathcal{N}\sqrt{-\gamma} \theta(\epsilon)^2,\quad
	L_f = \mathcal{N} \frac{5}{12} \sqrt{-\gamma} \theta(\epsilon)^4,
	\label{eq:conterterms}
\end{equation}
where $\gamma$ is the induced metric at $z=\epsilon$ near the AdS boundary and
$\sqrt{-\gamma} = \epsilon^{-4}$.
Substituting $\theta(u) = \theta_0 u + \theta_2 u^2 + \cdots$, we obtain
\begin{equation}
	L_1 = \frac{\mathcal{N}}{4} \frac{1}{\epsilon^4},\quad
	L_2 = - \mathcal{N} \left(
		\frac{1}{2}\frac{\theta_0^2}{\epsilon^2} + \theta_0 \theta_2
	\right),\quad
	L_f = \mathcal{N} \frac{5}{12} \theta_0^4.
\end{equation}
The coefficients $\theta_0$ and $\theta_2$ are related to the quark mass $m_q$ and the quark condensate $c$ by
\begin{equation}
	\theta_0 = m_q,\quad
	\theta_2 = c + \frac{1}{6}m_q^3,
	\label{eq:rel_mass_condensate}
\end{equation}
respectively.
We write
\begin{equation}
\begin{aligned}
	L_{ct} =& L_1 + L_2 + L_f
	=  \mathcal{N}\left[
		\frac{1}{4 \epsilon^4}
		- \frac{m_q^2}{2\epsilon^2}
		- m_q \left(c + \frac{1}{6}m_q^3\right)
		+ \frac{5}{12}m_q^4
	\right]\\
	=&
	\mathcal{N} \int_{u_h}^{\epsilon}\left[
		- \frac{1}{u^5} + \frac{m_q^2}{u^3}
	\right]\dd{u}
	+ \mathcal{N}\left(
		\frac{1}{4 u_h^4} - \frac{m_q^2}{2 u_h^2}
	\right)
	+ \mathcal{N} \left( \frac{m_q^4}{4} - m_q c \right).
\end{aligned}
\end{equation}
Then, the regularized free energy can be computed by
\begin{equation}
	F =\int_{u_h}^{\epsilon} \tilde{\mathcal{L}} \dd{u} + L_{ct}
	=
	\int_{u_h}^{\epsilon}\left[
		\tilde{\mathcal{L}}
		+ \mathcal{N}\left(
			- \frac{1}{u^5} + \frac{m_q^2}{u^3}
		\right)
	\right]\dd{u}
	+ \mathcal{N}\left(
		\frac{1}{4 u_h^4} - \frac{m_q^2}{2 u_h^2} + \frac{m_q^4}{4} - m_q c
	\right).
\end{equation}

In the multivalued region, $F(d)$ has swallowtail structure as a function of $d$, as shown in Fig.~\ref{fig:free_energy}.
In such cases, the intersection of the two branches of $F(d)$ is considered as a phase transition point.

The free energy is a thermodynamic potential in the canonical ensemble.
In this case, $d$ is treated as a controlling parameter.
On the other hand, we can also consider the grand canonical ensemble setup when we treat $\mu$ as a controlling parameter.  
In the grand canonical ensemble, the thermodynamic potential is given by the grand potential
\begin{equation}
	\Omega(\mu) = F - \mu d.
\end{equation}
Figure \ref{fig:grand_potential} shows $\Omega/T^4$ as a function of $\tilde{\mu}$ at $\tilde{T} = 0.343737$.
Note that there is the branch of the Minkowski embedding with vanishing density but finite $\mu$.
Considering both the Minkowski and BH embeddings, we can find swallowtail structure in $\Omega(\mu)$.
The intersection point of the swallowtail in $\Omega(\mu)$ is the first order phase transition point.
At low temperatures, the multivalued region of $\tilde{d}$ as a function of $\tilde{\mu}$ disappears.
Then, $\tilde{d}$ goes zero at finite $\tilde{\mu}$ without the multivaluedness.
It means the second order phase transition from the BH embedding to the Minkowski embedding.

The phase diagrams of the grand canonical and the canonical ensemble setups are shown in Fig.~\ref{fig:geo_phase_diagram}.

\bibliographystyle{utphys}
\bibliography{main}
\end{document}